\documentclass[%
 reprint,
 amsmath,amssymb,
 aps,
pra,
]{revtex4-2}
\usepackage{appendix}
\usepackage{comment}
\usepackage{amsmath}
\usepackage{amssymb}
\usepackage{bbold}
\usepackage{bm}
\usepackage{braket}
\usepackage[pagewise]{lineno}
\usepackage{csquotes}
\usepackage{derivative}
\usepackage{dcolumn}
\usepackage{dsfont}
\usepackage{comment}
\usepackage{graphicx}    
\usepackage{lineno}
\usepackage{mathtools}
\usepackage{slashed}
\usepackage{tikz}
\usepackage{verbatim}
\usepackage{xcolor}

\usepackage{booktabs}
\usepackage{nicematrix} 
\definecolor{giallo}{RGB}{255, 255, 0}
\definecolor{rosso}{RGB}{255, 0, 0}
\definecolor{verde}{RGB}{0, 128, 0}
\definecolor{blu}{RGB}{0, 0, 255}
\definecolor{white}{RGB}{255, 255, 255}
\usepackage{braket}
\begin{document}

\preprint{APS/123-QED}


\title{Single particle dynamical signature of topology induced by single mode cavities in Su-Schrieffer-Heeger chain}

\author{F. Pavan$^{1,*}$}\author{G. Di Bello$^{1,2}$}\author{G. De Filippis$^{2,3}$}\author{C. A. Perroni$^{2,3}$}
\affiliation{$^{1}$Dip. di Fisica E. Pancini - Università di Napoli Federico II - I-80126 Napoli, Italy}
\affiliation{$^{2}$INFN, Sezione di Napoli - Complesso Universitario di Monte S. Angelo - I-80126 Napoli, Italy}
\affiliation{$^{3}$SPIN-CNR and Dip. di Fisica E. Pancini - Università di Napoli Federico II - I-80126 Napoli, Italy}
\affiliation{$^*$Corresponding author: F. Pavan, fabrizio.pavan@unina.it }

\date{\today}

\begin{abstract}
\color{black}
Witnessing and tracking topological phase transitions induced by interactions with the environment is a crucial challenge. Among the various experimental approaches to detect topological properties, the Mean Chiral Displacement (MCD) has emerged as a powerful bulk probe in one-dimensional chiral systems, allowing the extraction of the topological invariant from single-particle dynamics. Here we study the dynamics of a single particle in a one-dimensional Su–Schrieffer–Heeger chain coupled to multiple cavity modes via inter-cell hopping terms, focusing on the out of equilibrium behavior of the MCD. We show that, whenever the frequency is larger than the static hopping amplitudes, the coupling induces a discontinuous jump in the MCD, already at small times, signaling that such a coupling also leaves a signature in the survival edge probability when the dynamics are initialized at one of the two edges. For frequencies comparable to the static hopping amplitudes, topological order competes with dissipative effects, which makes the MCD behaves smoothly, retaining information about the driven-dissipative topology. 

\color{black}

\end{abstract}

\maketitle

\section{Introduction}

Since the discovery of the Quantum Hall effect, topological phases have occupied a central position in condensed matter physics owing to their unconventional physical properties \cite{QHE}. Their description lies beyond the conventional Ginzburg-Landau paradigm: rather than emerging from a symmetry-breaking mechanism or being characterized by local or long-range order parameters, they are defined through a global topological order \cite{PhysRevA.110.023318,PhysRevLett.49.405,PhysRevLett.50.1153,PhysRevLett.61.2015,RevModPhys.82.3045,Bernevig2013}. This distinctive feature is encoded in the discontinuous behavior of bulk topological invariants, defined under periodic boundary conditions (PBCs). As a result, observables such as transport coefficients also display discontinuous changes. Whenever the topological invariant takes a non-zero value, it signals the emergence of edge-localized states in the corresponding system under open boundary conditions (OBCs). This universal property, shared by all topological systems, embodies the principle known as bulk-boundary correspondence.  

Experimentally, the most common approach to probing non-trivial topology is the detection of localized boundary states \cite{RevModPhys.91.015006}. Measuring topological invariants directly is also possible, but it generally represents a demanding task, since it requires access to the bulk eigenfunctions \cite{Atala2013-bq}. In many experimental platforms this is either extremely challenging or even not feasible, particularly when the system boundaries are not easily accessible.  
It is widely known that symmetries play a pivotal role in matter of classification of topological phases and their properties and mechanism protection.
In particular, for one dimensional systems where the edge states are protected by the chiral symmetry, an alternative route to detect topological properties was introduced in \cite{Cardano2017}. In this work focus on the dynamics of a single particle in 1D chiral systems and on the behavior of a  suitable bulk observable, the Mean Chiral Displacement (MCD). They showed that, by exploing single-particle dynamics in chiral lattices, the MCD converges to the topological invariant in the long-time limit. This method is advantageous because does not require external fields, and provides a viable strategy to probe topology even in chiral systems without ill-defined or inaccessible edges, by exploiting single-particle dynamics. The MCD has been successfully implemented in photonic quantum walks to detect topological phases \cite{Cardano2017,Maffei_2018,CardanoQuench}, by mean of Bose-Einstein condensates \cite{MCD-bose1}, and in acoustic systems in the presence of disorder \cite{MCD-acoustic}.

Up to now, the MCD has been mainly considered in fermionic systems, and, in most cases, in the non-interacting case. More recently, theoretical works have analyzed its robustness in the presence of fermionic interactions \cite{MassignanRizzi} and for driven-dissipative systems \cite{MCD-photonic}. The current literature focuses on the robustness of MCD with respect to interactions, and most efforts have concentrated on assessing its stability across different coupling regimes. It remains unclear, however, whether the MCD can be used as a marker of interaction-induced topological phase transition. 

At the same time, a complementary perspective has emerged in which dissipation is not only detrimental but may also act as a resource to induce topological phase transitions. 
Moreover, an additional standpoint arises when considering systems coupled to
external environments. By tracing out the environmental degrees of freedom, the
system’s dynamics become non-unitary and can be effectively described by a non-Hermitian Hamiltonian. From a topological condensed matter perspective, the inclusion of non-Hermiticity greatly enriches the physical landscape. Indeed,
the Altland–Zirnbauer tenfold way, which classifies Hermitian topological phases,
expands dramatically in the non-Hermitian regime, giving rise to as many as 38
symmetry-protected topological classes ~\cite{NH1,NH2,NH3,Ashida2020_NonHermitianReview}. 
As a consequence, several attempts have been made to extend the definition of the MCD to non-Hermitian systems \cite{MCD-NH1,MCD-NH2}.

Following this idea, several works have explored the coupling between different quantum fields can drive interaction-induced topological transitions \cite{Schiro2022, Schiro2024, Segal,Nori-dissipative-topo}. In particular, in a recent work of some of us \cite{Pavan}, we showed that a Su-Schrieffer-Heeger (SSH) chain at half filling, coupled to many local environments through hopping amplitudes, can undergo topological phase transitions. Specifically, we demonstrated that in one version of the model, the interaction turns a topological phase into a trivial one, while in another version it converts a trivial phase into a topological one. In both cases, the coupling to the baths induces a topological phase transition, which we analyzed in detail under PBCs and OBCs. However, in our previous work we focused only on static properties, without discussing dynamical signatures of this transition or how it could be experimentally accessed.

Motivated by these reasons, in the present work, we analyze the 
Su-Schrieffer-Heeger chain where one hopping particle is affected by the coupling to  single mode cavities \cite{perroni1,perroni2,perroni3}. We identify signatures of topology induced by such interaction effects already at the single-particle level, by analyzing the dynamics of a wave packet, initially localized in the central unit cell and in conditions where an experimental setup can easily be realized. We will show that, in the anti-adiabatic limit, the MCD reveals the role of external coupling as a driver of interaction-induced topological transitions, now manifested in the single-particle dynamics. This picture is further corroborated by studying the survival edge probability, which attains a finite value whenever the MCD oscillates around 0.5. By decreasing the interaction strength and making the cavity frequency comparable to the intra-cell and inter-cell hopping amplitudes, we observe that the MCD exhibits a continuum-like behavior, signaling a competition between dissipative effects. 
The paper is organized as follows. In Sec. \ref{section 2} we introduce the SSH model and review the single-particle dynamical probes of topology, with particular emphasis on the MCD and the survival edge probability.
In Sec. \ref{section 3} we present the open SSH model where the chain is coupled to cavity modes, and we characterize the interaction-induced topological signatures emerging in the single-particle dynamics.
In Sec. \ref{section 4} we analyze the role of the cavity frequency, studying the dependence of MCD and surivival edge from it. Finally, in Sec.~\ref{section 5} we analyze the regime in which the cavity frequency is comparable to the static electronic hopping amplitudes. Although reaching the long-time limit is numerically demanding and therefore not accessible within our simulations, we observe that the interaction induces a distinct polarization between the two sublattices in the left and right sides of the chain.

\section{Closed SSH model} \label{section 2}
\begin{figure}[htbp!] 
    \begin{center}   
        \includegraphics[scale=0.5]{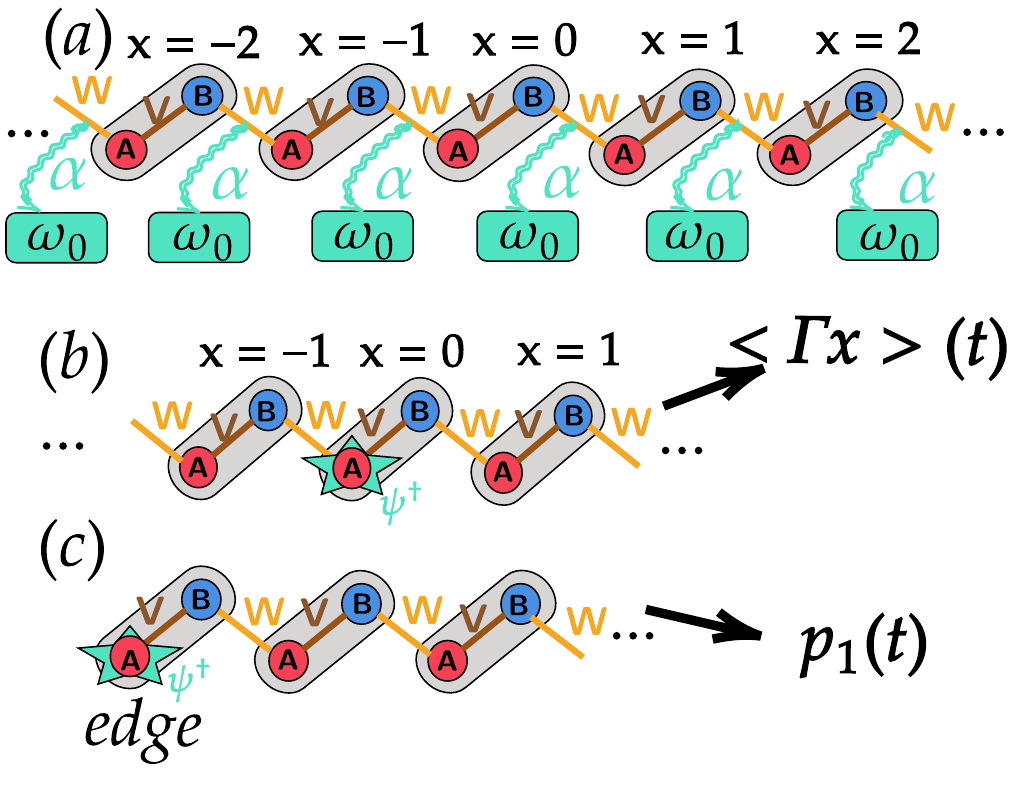} \caption{\label{fig1} Panel (a): Open SSH chain coupled to inter-cell hopping with several cavity modes at fixed frequency $\omega_0$. The central unit cell is set as the origin of the position reference frame.
Panel (b): The dynamics used for the MCD are initialized from a state localized in the central unit cell.
Panel (c): The survival edge probability is evaluated for dynamics starting with the electron initially localized at one of the two edges.
}
    \end{center} 
\end{figure}
The SSH model (see Fig.\,\ref{fig1}(a)) represents a paradigmatic toy model of a one-dimensional topological insulator. The single electron hops on a 1D chain that consists of $N$ unit cells, each hosting two electronic sites belonging to different sublattices, indicated as $A$ and $B$. The intra-cell hopping is $v$, while the inter-cell hopping is $w$. The entire chain is therefore built from a total of $2N$ electronic sites with staggered hopping amplitudes, described by the following tight-binding Hamiltonian:
\begin{equation} \label{bare SSH}
  H_{SSH}=(v\sum^{N}_{n=1}c^{\dagger}_{n,A}c_{n,B} +w\sum_{n=1}^{N-1} c^{\dagger}_{n+1,A}c_{n,B})+h.c.  \,\, .
\end{equation}
where $c_{n,\nu}$ ($c^{\dagger}_{n,\nu}$) destroys (creates) an electron in the site $\nu=A,B$ of the $n$-th cell, with $n=1,...,N$, and $v$ ($w$) denotes the intra- (inter-) cell hopping. With OBC and for $v>w$, the model is in the trivial insulating phase, in which electrons are delocalized in the bulk. On the contrary for $v<w$, the model is in the topological phase, where localized gapless edge states appear at the ends of the chain. From a parallel point of view, by considering the SSH chain under PBC, we can express $H_{SSH}=\sum_{k} H(k)$, where:
 \begin{equation}
     H(k)= E_k  \, \mathbf{n}(k) \cdot  \boldsymbol{\sigma} ,
 \end{equation}
with $E_k=\sqrt{v^2+w^2+2vw \cos{k}}$ is the band dispersion, $\mathbf{n}(k)=((v+w \cos{k})/E_k,(w \sin k)/E_k , 0)$ and $\boldsymbol{\sigma}$ is the vector of Pauli matrices. Under PBC, it is possible to define a bulk quantity called winding number:
\begin{equation}
    \eta=\frac{1}{2\pi} \int_{-\pi}^{\pi} dk (n_{x}(k)\partial_k n_{y}(k)-n_{y}(k)\partial_k n_{x}(k)) .
\end{equation}
This quantity plays the role of topological invariant, assuming the value 0 in the trivial phase (meaning $v>w$) and 1 in the topological phase (for $v<w$).
Thus, the non-zero value of the topological invariant is directly related to the number of edge states. This mechanism, arising in this toy model, is fully general and is well-known in literature as bulk-boundary correspondence. 
For 1D chiral system the topological invariant is closely related to the mean chiral displacement, defined as:
\begin{equation} \label{def MCD}
    \langle{\Gamma x}\rangle (t)=\sum_{\nu} 
\left\langle c_{0,\nu} e^{i H t} \Gamma x   e^{-i H t} c^{\dagger}_{0,\nu}  \right\rangle ,
\end{equation}
where $\Gamma$ is the chiral symmetry operator, which anticommutes with the Bloch Hamiltonian if the symmetry is present, and $x$ is the position operator, i.e. $\{\Gamma,H\}=0$ and $\Gamma^2=1$. By creating an electron in the central unit cell, which is defined as the origin of the position along the chain ($x=0$) (see Fig.\,\ref{fig1}(b)), it has been demonstrated that the MCD converges to the topological invariant in the long-time limit \cite{Cardano2017,MassignanRizzi,Maffei_2018}. As a consequence, for $v>w$, the MCD in Eq.\,\eqref{def MCD} oscillates around $0$, whereas for $v<w$ it oscillates around $0.5$.

In the case of the bare SSH model the Eq.\,\eqref{def MCD} assumes the following form \cite{Cardano2017,MassignanRizzi}:
\begin{multline}
    \langle{\Gamma x}\rangle(t)= \frac{\eta}{2} \\  -\int_{-\pi}^{\pi} \frac{dk}{4 \pi i}  \cos({2 E_k t}) (n_{x}(k)\partial_k n_{y}(k)-n_{y}(k)\partial_k n_{x}(k)) ,
\end{multline}
where the first term is proportional to the topological invariant, while the second term gives rise to oscillations modulated by the energy band $E_k$.

\begin{figure}[htbp!] 
\centering
        \includegraphics[scale=0.18]{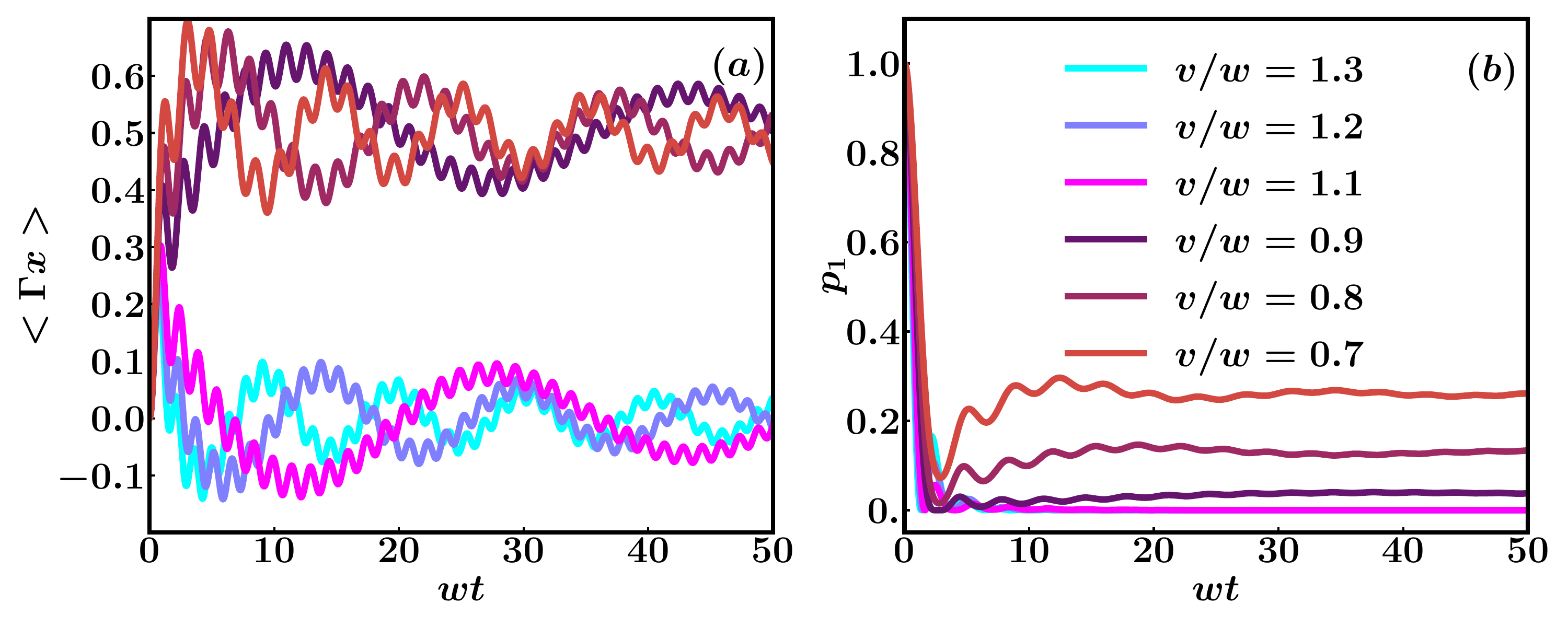} \caption{\label{fig2} Mean Chiral Displacement (panel (a)) and survival edge probability (panel (b)) for the bare SSH model (described by Eq.\,\eqref{bare SSH}) as functions of different ratios of the hopping $v/w$.}
\end{figure}

Thus, by exploiting single-particle dynamics, the MCD converges to the topological invariant in the long-time limit (see Fig. \ref{fig2}(a) where we plot the MCD for different hopping ratio $v/w$). 
As stressed above, this is particularly useful to detect topological properties in systems where the boundaries are not fully accessible from an experimental point of view, without having access to edge states directly.

From a complementary perspective, and focusing on the case of OBC, we can also consider dynamics starting from one of the two edges. By focusing again on the bare SSH model, we can define the survival edge probability as $p_1(t)=|\langle A_1|e^{-iHt}|A_1\rangle|^2$, where $\ket{A_1}$ is the initial state with the electron localized at one of the two edges of the SSH chain (in our simulations, the left edge, as shown in Fig.\,\ref{fig1}(c)) and $\ket{\psi(t)}=e^{-iHt}\ket{A_1}$. When $v>w$, this quantity starts from 1 and rapidly decays to zero. In the opposite case, $v<w$, this probability converges to a non-zero value, indicating that signatures of formation of edge states are present even at the level of single-particle dynamics. 
In the Appendix \ref{appendix probability}, we have that the overlap $\braket{A_1|\psi(t)}=\braket{A_1|e^{-iHt}|A_1}$ is given by:

\begin{equation}\begin{split}\label{At}
&\mathcal A(t) = \braket{A_1|e^{-iHt}|A_1}
= \Theta(w-v)\Big(1-\frac{v^2}{w^2}\Big)
 \\ &+\frac{2}{\pi}\int_0^\pi \mathrm dk\;
\frac{v^2\sin^2 k}{\,v^2+w^2+2vw\cos k\,}\;
\cos(E_k t),
\end{split}\end{equation}
where the first term is the discrete edge contribution present only for $v<w$.
The survival edge probability is then given by $p_1(t)=| \mathcal A(t)|^2$ and its long-time limit is:
\begin{equation}
\lim_{t\to\infty}p_{1}(t)=
\begin{cases}
\big(1-\frac{v^2}{w^2}\big)^2, & v<w,\\[2pt]
0, & v\ge w.
\end{cases} \;\; ,
\end{equation}
which is non-zero only in the topological phase $w>v$. Moreover, by lloking at Fig. \ref{fig2},for ratios $v/w<1$ such a probability reaches non-zero stationary values.
This general picture is rather simple, yet at the same time it reveals the presence of bulk–boundary correspondence even by looking at single-particle dynamics in the bare SSH model.
\section{Open SSH model} \label{section 3}
\begin{figure}[htbp!] 
    \begin{center}   
        \includegraphics[width=1.05\linewidth]{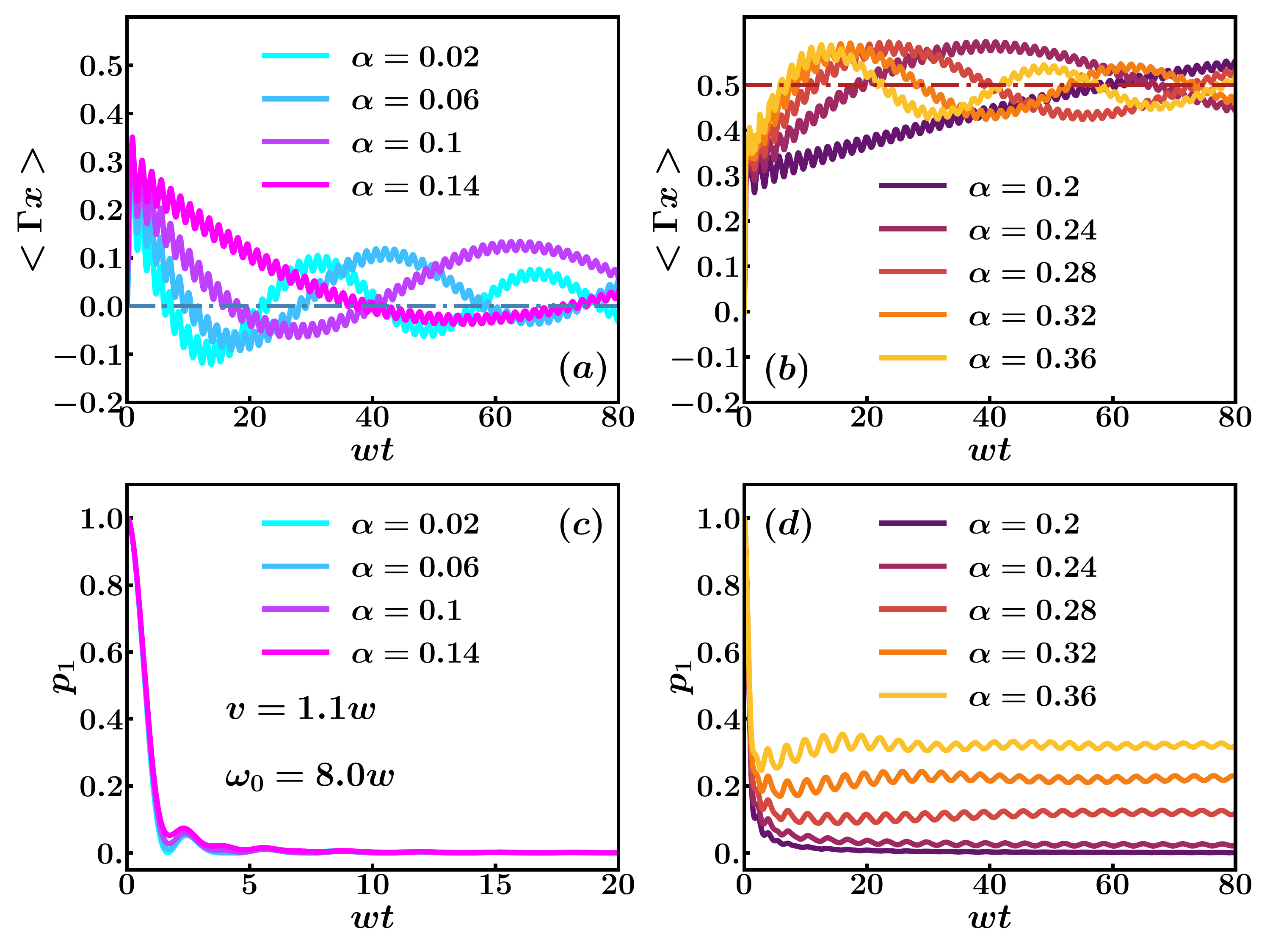} \caption{\label{fig3} Mean Chiral Displacement (panels (a) and (b)) and survival edge probability (panels (c) and (d)) for the SSH model as functions of dimensionless time, for different values of the coupling $\alpha$ and for $N=201$ unit cells.}
    \end{center} 
\end{figure}

Following the spirit of \cite{Pavan}, we will show that a similar behavior can be induced by an external coupling to several cavity modes. To this end, we perform numerical simulations based on a tensor network approach, by mean of ITensor library in \textit{Julia} \cite{fishman2022itensor}, focusing on single-particle dynamics governed by the Hamiltonian $H=H_{SSH}+H_B+H_{SSH-B}$, where $H_B=\omega_0\sum_n b^\dagger_n b_n$ is the collection of local harmonic oscillators of fixed frequency $\omega_0$, and $H_{SSH-B}$ represents the interaction term,
\begin{equation} \label{interaction hamiltonian}
    H_{SSH-B}=g\sum_{n=1}^{N-1} (b^\dagger_n+b_n) (c^\dagger_{n+1,A}c_{n,B}+h.c.),
\end{equation}
where $b_n$($b^\dagger_n$) is the operator that annihilates(creates) an excitation in the $n$-th cavity. From now on, we introduce the dimensionless parameter $\alpha$, defined by $g=\sqrt{\alpha/2} \omega_0$ measuring the strength of the coupling. 
The dynamics was simulated starting from two initial conditions: one where the electron is localized on sublattice A of the central unit cell, used for the results on the MCD, and another where the electron is localized on sublattice A of the first unit cell, used for the results on the survival edge probability. The bosonic oscillators were initialized in the vacuum state, and the time evolution was performed using the $W^{I}$ algorithm with complex time steps \cite{paeckel2019time,wII-Polman}, i.e., by applying to the state a matrix product operator constructed from the second-order expansion of the evolution operator. For such a system, it is not possible to use the conventional Time-Dependent Variational Principle (TDVP), since the interaction term given by Eq.\,\eqref{interaction hamiltonian}, that we are going to consider, is expressed as sum of products of three operator (see Appendix for details). 
The Matrix Product State (MPS) chain is built by alternating two fermionic and one bosonic site (see Appendix \ref{convergence section}); this structure ensures the bond dimension does not grow excessively.
Fig.\,\ref{fig3} shows the results obtained for version (b) of the model in Fig.\,\ref{fig1}.
We start from the situation where $v/w=1.1$ and the frequency of the bosonic oscillators is set to $\omega_0=8 w$.    

Panels (a) and (b) of Fig.\,\ref{fig3} show the behavior of the MCD, defined in Eq.\,\eqref{def MCD}, as a function of the boson–fermion coupling $\alpha$, starting from an electron localized on sublattice A of the central unit cell.
In the weak–coupling regime, illustrated in panel (a), the MCD oscillates around zero. 
However, one can observe that the frequency of the global oscillation varies with $\alpha$, indicating that the interaction drives the closing of the energy gap. Indeed, in this weak-coupling regime, as the interaction increases, the frequency of the oscillations around the topological invariant decreases.
The curves in panel (b) begin to converge towards a non-trivial topological invariant for $\alpha \geq 0.2$. In this regime, the oscillation frequency increases with increasing interaction, and the topological invariant fluctuates around 0.5 with more pronounced global oscillations.
Information about the presence of an induced topology is also encoded in the survival edge probability. As shown in panels (c) and (d) of Fig.\,\ref{fig3}, for couplings $\alpha < 0.2$ the survival edge probability rapidly decays to zero, almost independently of the boson–fermion interaction strength. In contrast, for stronger interactions the survival edge probability approaches stationary values different from zero, consistent with the behavior displayed by the MCD. 

\begin{figure}[htbp!] 
    \begin{center}   
        \includegraphics[scale=0.20]{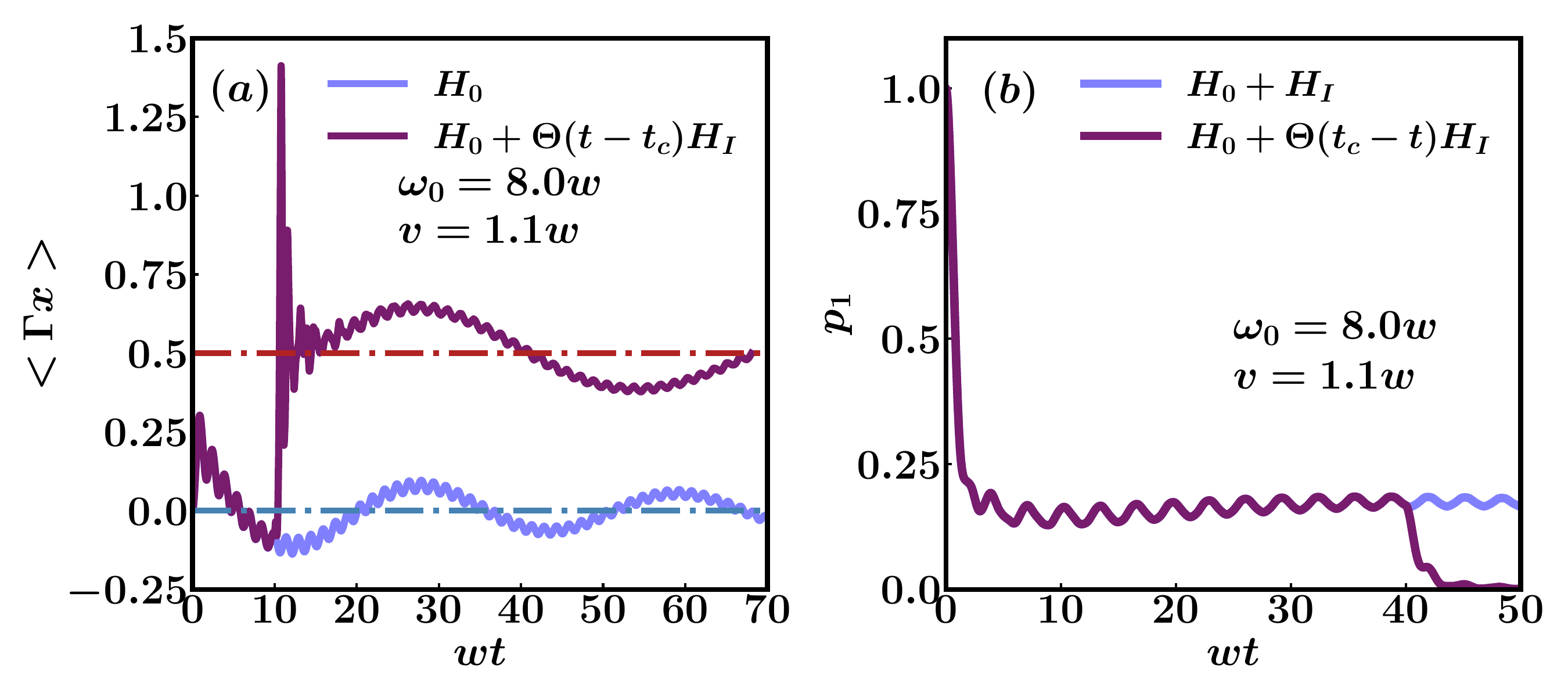} \caption{\label{fig4} Mean Chiral Displacement (panel (a)) and survival edge probability (panel (b)) as function of time after a sudden quench. Purple curves correspond to $\alpha=0.3$.}
    \end{center} 
\end{figure}

In Ref.\,\cite{CardanoQuench} it is pointed out that the MCD signals the emergence of topology also performing a sudden quench between two different Hamiltonians. 
In order to further support our findings, we considered a sudden quench implemented by switching on or off the interaction term given by Eq.\,\eqref{interaction hamiltonian}. Fig.\,\ref{fig4} shows the sudden quench dynamics for the version (a) of the model of Fig.\,\ref{fig1}. Fig.\,\ref{fig4}(a) shows the dynamics of single electron starting from the central unit cell. In the early-time dynamics ($wt<10$), the evolution is driven by the bare SSH Hamiltonian given by Eq.\,\eqref{bare SSH}, then the interaction term is suddenly switched on. Consequently, the MCD shows a discontinuous behavior at the quench time. Then, going far away from the quench time the MCD oscillates around a non-trivial value of the topological invariant.
In Fig.\,\ref{fig4}(b) the survival edge probability is shown for $\alpha=0.3$ and $\omega_0=8w$ (blue curve), together with its variant dynamics (purple curve) when the interaction term is suddenly switched off for $wt>40$. In the latter case, the electron rapidly leaves the edge, further stressing that the formation of the edge state is induced by the interaction.

\section{Cavity Frequency Behavior} \label{section 4}
\begin{figure}[htbp!] 
    \begin{center}   
        \includegraphics[scale=0.2]{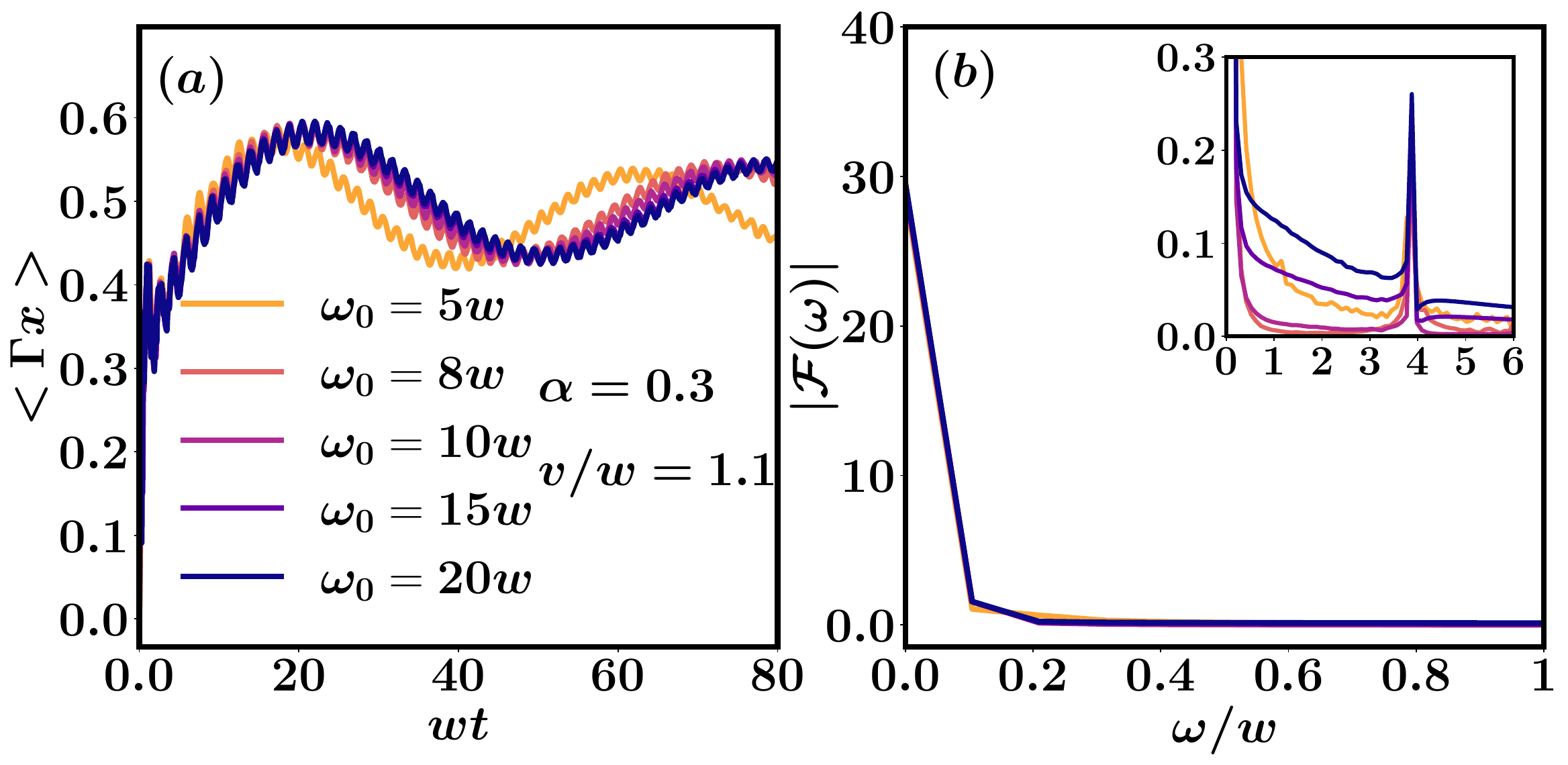} \caption{\label{fig5} Mean Chiral Displacement (panel (a)) and its FFT (panel (b))  as functions of dimensionless time and frequency respectively, for different values of the frequency $\omega_0$ at coupling $\alpha=0.3$.}
    \end{center} 
\end{figure}

\begin{figure}[htbp!] 
    \begin{center}   
        \includegraphics[scale=0.2]{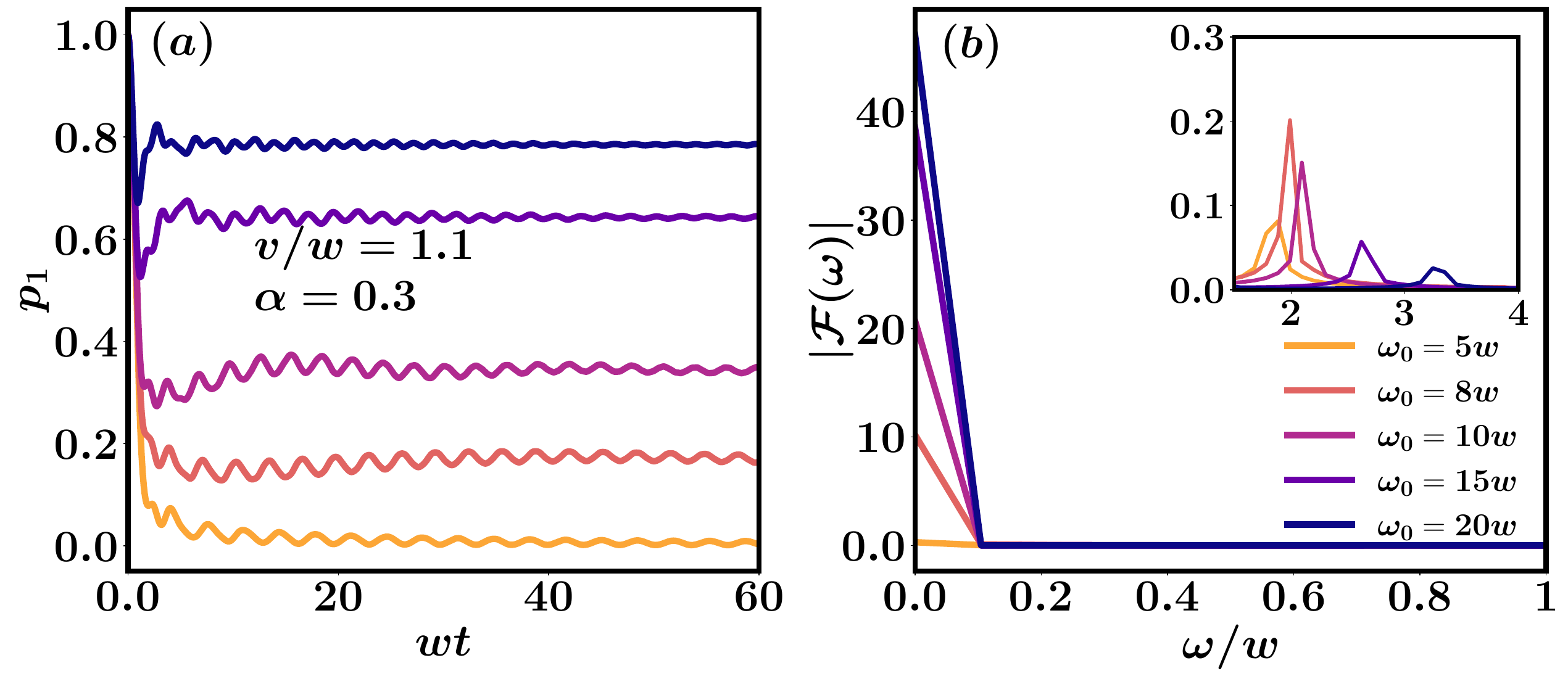} \caption{\label{fig6}  Survival edge probability (panel (a)) and its FFT (panel (b)) as functions of dimensionless time and frequency, respectively, for different values of the boson frequency $\omega_0$ at coupling $\alpha=0.3$.}
        \color{black}
    \end{center} 
\end{figure}

By examining Fig.\,\ref{fig2}, it appears that the bulk–boundary correspondence is preserved even in the presence of boson–fermion coupling. However, when considering different boson frequencies, this correspondence can break down. In particular, in the high-frequency limit corresponding to the anti-adiabatic regime, the survival edge probability reaches non-zero stationary values for coupling strengths lower than those at which the MCD exhibits a sharp jump. As we move toward a regime where the frequency $\omega_0$ approaches the values of the hopping amplitudes $v$ and $w$, while still satisfying $\omega_0 > v,w$, the induced topology is signaled earlier by the MCD. However, it is important to emphasize that for all frequencies $\omega_0 > v, w$, signatures of interaction-induced topology are present in both quantities. 
Fig.\,\ref{fig5}(a) shows the MCD at a fixed value of the coupling parameter, $\alpha = 0.3$, for different bosonic frequencies. We observe that the MCD is almost independent of the boson frequency, displaying the same physical behavior across all curves. This picture is further confirmed by the Fast Fourier Transform (FFT), shown in Fig.\,\ref{fig5}(b) and in the corresponding inset. The behavior at low frequencies is identical, clearly indicating that all curves oscillate around the non-trivial value of the topological invariant, namely 0.5. Moreover, the inset shows that all curves share almost the same bulk oscillation frequency.
This behavior changes dramatically when we consider numerical simulations starting from the left edge. The results are reported in Fig.\,\ref{fig6}. Panel (a) shows the survival probability for different frequencies, again evaluated for $\alpha = 0.3$. We observe that the higher the frequency, the higher the probability associated with the survival of the left edge. In particular, the probability decays to zero for $\omega_0 = 5w$, while at higher frequencies it tends toward distinct finite values. In the high-frequency limit, the external coupling effectively confines the electron, keeping it highly localized at the edge. This is evident in panel (b): the main peak of the FFT becomes more pronounced near zero frequency, indicating a higher stationary value of the survival probability as the cavity frequency increases. Furthermore, the inset of Fig. \ref{fig6} reveals a systematic increase in the oscillation frequency around the stationary probability as the cavity frequency is increased.

\begin{figure}
    \centering
    \includegraphics[width=1.05\linewidth]{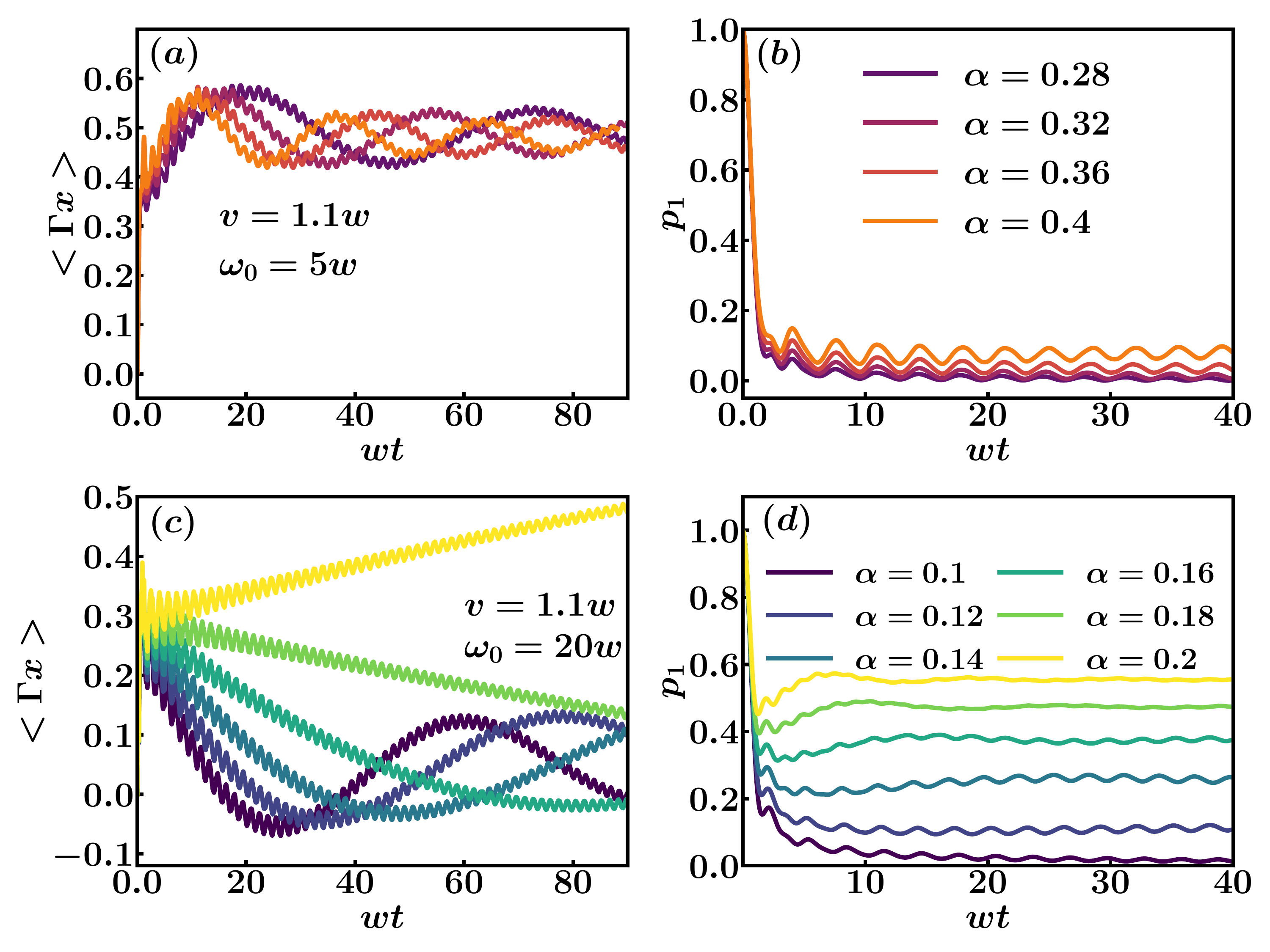}
    \caption{Panels (a) and (b): MCD and survival edge probability as functions of time, for different couplings $\alpha$ and for $\omega_0=5.0w$. Panels (c) and (d): MCD and survival edge probability as functions of time, for different couplings $\alpha$ and for $\omega_0=20.0w$. All figures are computed for $N=201$ unit cells.}
    \label{fig7}
\end{figure}

Overall, this analysis suggests that the two types of dynamics experience different renormalizations of the intra-cell hopping amplitudes induced by the interaction with the bosons. This observation is further supported by a closer comparison between the two limiting cases, $\omega_0 = 5w$ and $\omega_0 = 20w$.
Fig.\,\ref{fig7} illustrates these opposite behaviors. For $\omega_0 = 5w$, the MCD, shown in Fig. \,\ref{fig7}(a), indicates the onset of a coupling-induced topological phase before the survival edge probability shows any response ( as illustrated in Fig. \,\ref{fig7}(b), even though the MCD has already jumped to a non-trivial value. In Fig.\,\ref{fig7}(c) and (d), we plot the MCD and the survival edge probability in the high-frequency limit ($\omega_0 = 20w$), respectively. 
Conversely, the survival edge probability reaches non-zero stationary values before the MCD exhibits a jump to the topological invariant.

\section{Approaching the $\omega_0 \simeq v,w$ regime} \label{section 5}
When the MCD oscillates around a value different from zero, signalling the emergence of a topological phase, this indicates that propagation to the right and to the left of the central unit cell (chosen as the origin of the position reference frame) is no longer symmetric between the two sublattices \(A\) and \(B\). This asymmetry reveals the appearance of topological features. In order to prove this statement, we introduce the following decomposition of the MCD $\braket{\Gamma x}=\langle x_{R,A} \rangle-\langle x_{R,B} \rangle + \langle x_{L,A} \rangle- \langle x_{L,B} \rangle$, with:

\begin{equation}\label{right}
\langle x_{R,A/B} \rangle = \sum_{i=0}^{N/2} i\, \langle c^\dagger_{i,A/B} c_{i,A/B} \rangle,
\end{equation}
\begin{equation}\label{left}
\langle x_{L,A/B} \rangle = \sum_{i=-N/2}^{0} i\, \langle c^\dagger_{i,A/B} c_{i,A/B} \rangle,
\end{equation}
where \(\langle x_{R,A/B} \rangle\) and \( \langle x_{L,A/B} \rangle\) represent the contributions to the position of the electron on the right and left sides of the chain, respectively, projected onto sublattice \(A\) or \(B\).

Focusing on the right part of the chain, the electron moves from sublattice \(B\) to sublattice \(A\) through the inter-cell hopping \(w\). Conversely, to move from sublattice \(A\) to sublattice \(B\), the electron must hop through the intra-cell hopping \(v\). For this reason, in the topological phase where \(w>v\), propagation on sublattice \(A\) is faster than on sublattice \(B\). When focusing on the left part of the chain, the opposite holds. Namely, propagation on sublattice \(B\) is favoured over that on sublattice \(A\). In other words, although the overall polarization of the entire chain remains zero, in the topological phase the left and right parts exhibit opposite local polarizations.

\begin{figure} \centering \includegraphics[width=1.15\linewidth]{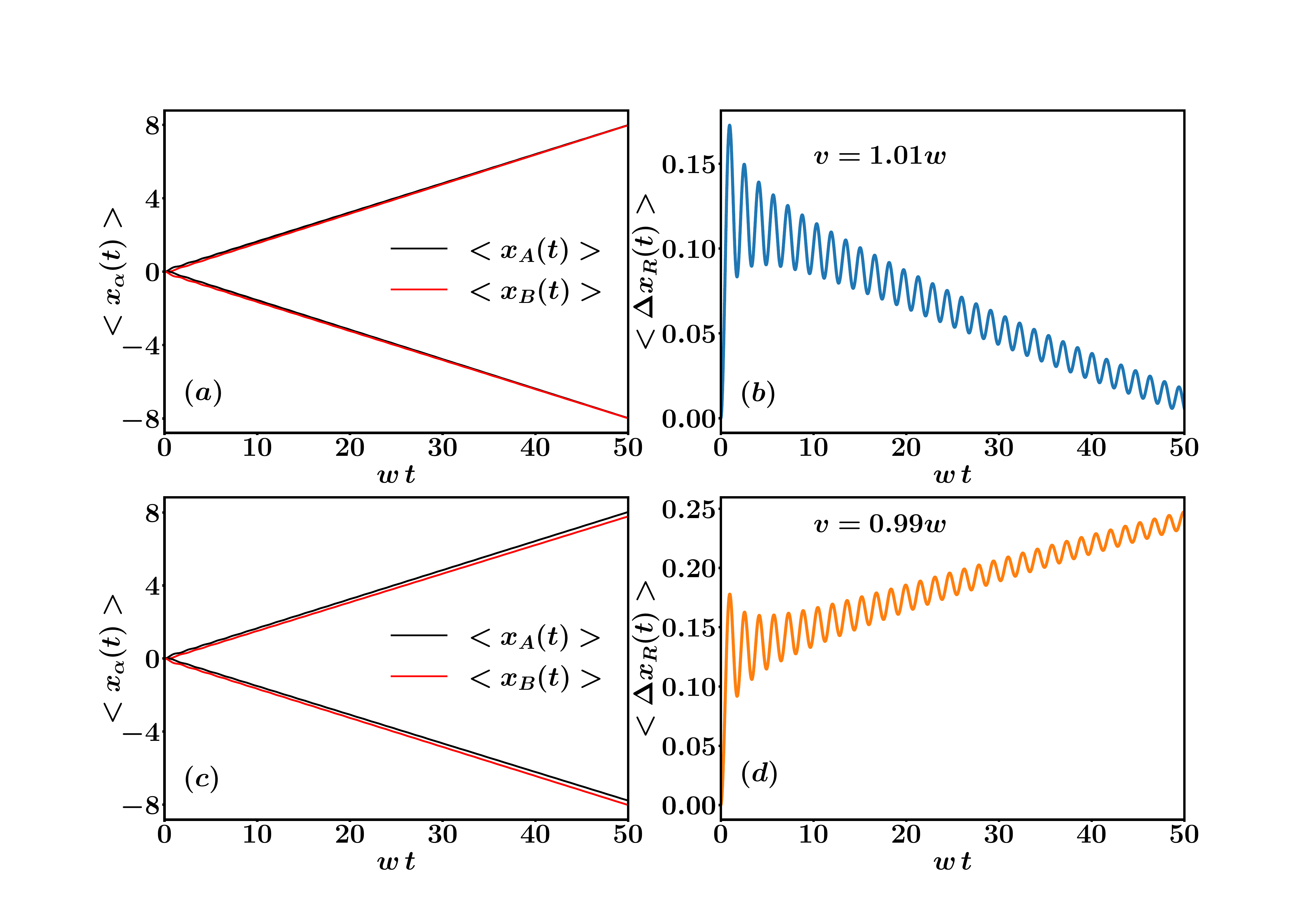} \caption{Left and right propagation within the sublattices as a function of dimensionless time, and difference between the two sublattice $A$ and $B$ rightward propagation of the electron, for $v/w=1.01$ (panels (a) and (b)) and $v/w=0.99$ (panels (c) and (d)) as a function of dimensionless time. } \label{bare MCD dec} \end{figure}

\begin{figure} \centering \includegraphics[width=1.05\linewidth]{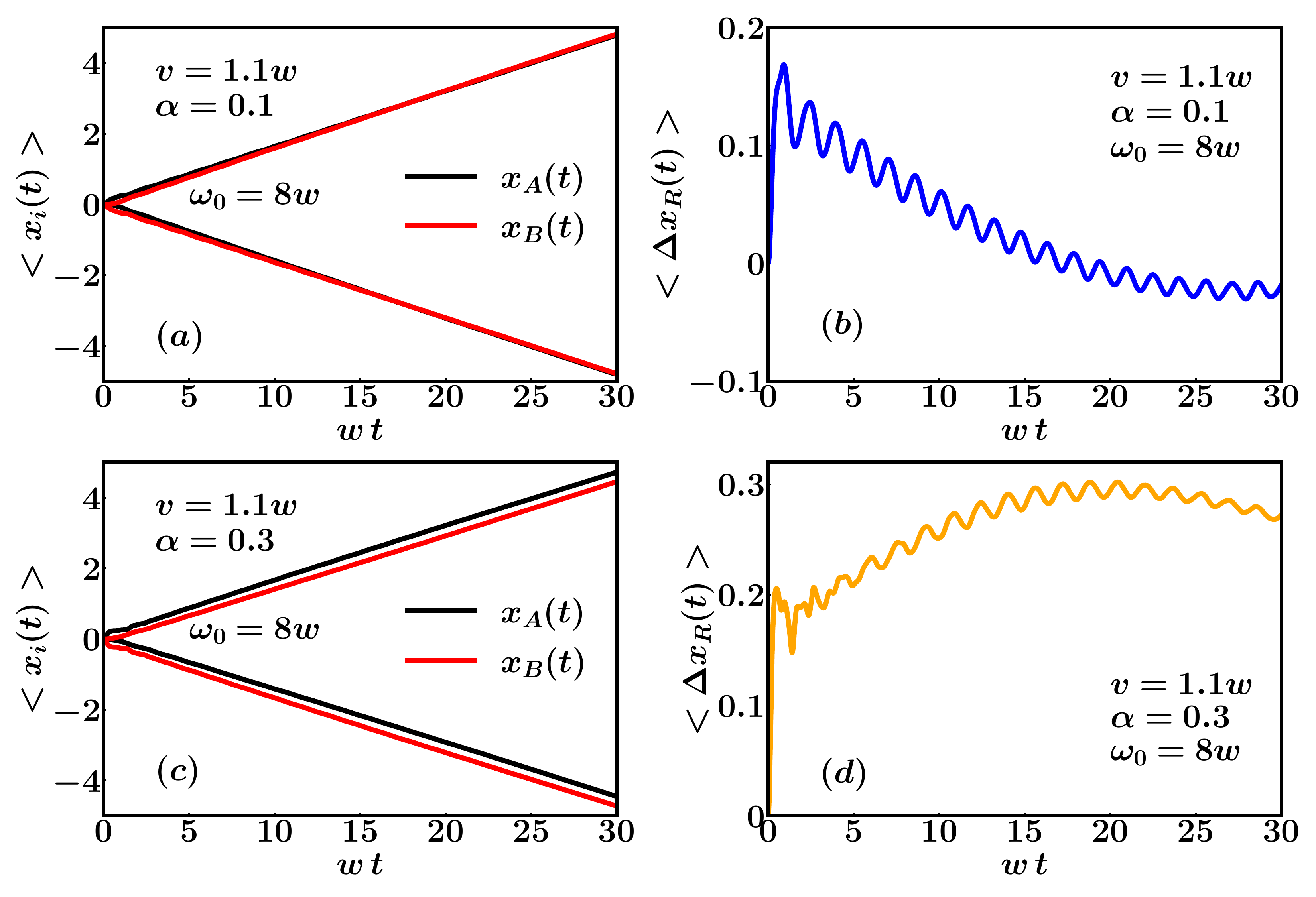} \caption{Panel (a) (Panel (c)): decomposed components of the MCD  for $\omega_0=8.0w$ and $\alpha = 0.1$ ($\alpha= 0.3$). Panel (b) (Panel (d)): difference in the rightward propagation for $\omega_0=8.0w$ and $\alpha = 0.1$ ($\alpha= 0.3$).} \label{omc8 dec} \end{figure}

\begin{figure} \centering \includegraphics[width=1.0\linewidth]{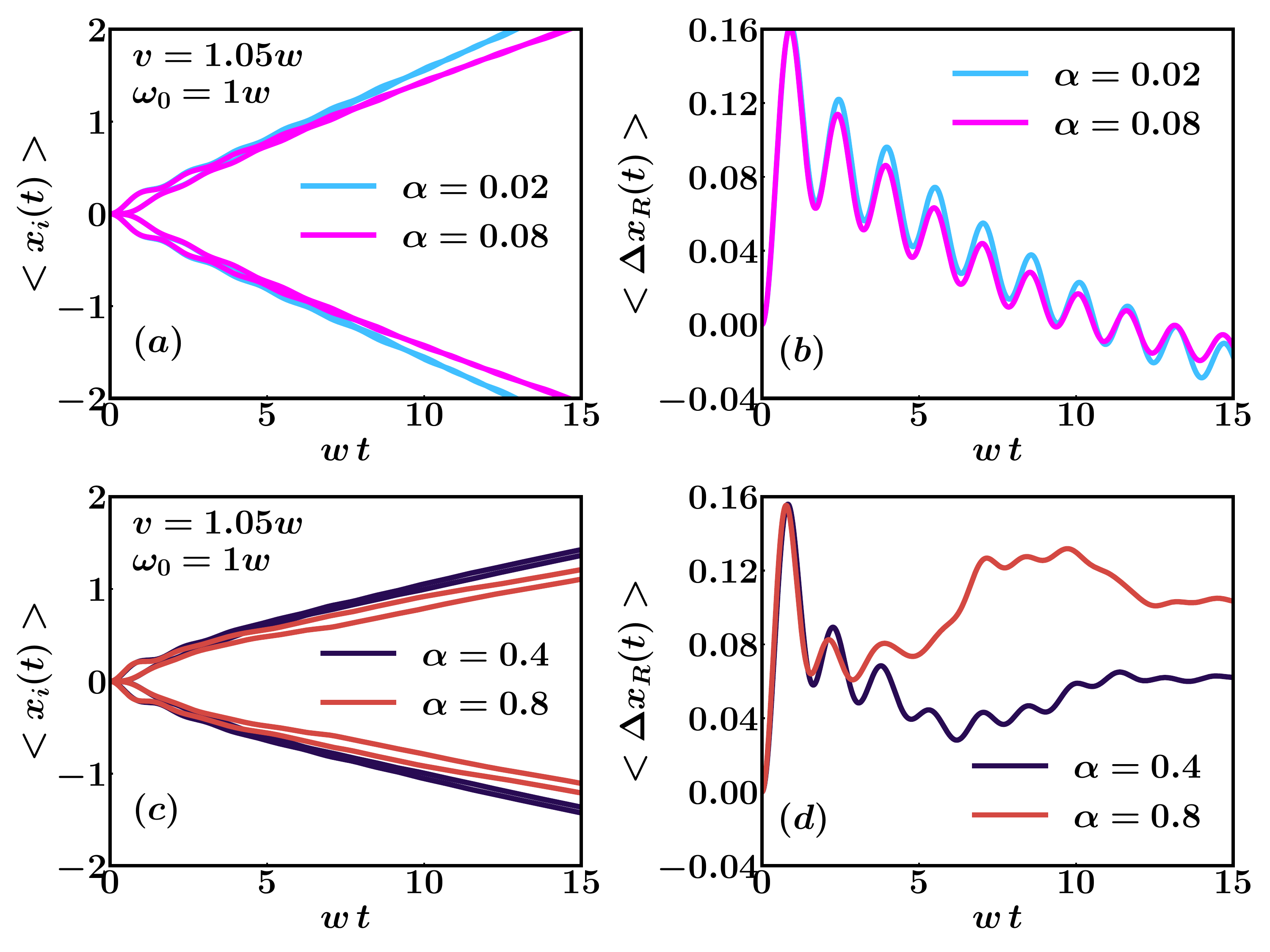} \caption{Panel (a) (Panel (c)): decomposed components of the MCD  for $\omega_0=1.0w$ in the weak(strong) coupling regime. Panel (b) (Panel (d)): difference in the rightward propagation for $\omega_0=1.0w$ in the weak(strong) coupling regime. } \label{omc1 dec} \end{figure}

In Fig.\,\ref{bare MCD dec} we show the bare case for the decomposed components of the MCD. To start from a balanced initial condition, the dynamics are computed beginning from the state
\begin{equation}
\lvert \psi_0 \rangle = \frac{1}{\sqrt{2}}\left( c^\dagger_{x=0,A} + c^\dagger_{x=0,B} \right)\ket{0}_{el},
\end{equation}
which is an equal superposition of the two sublattices localized at the central unit cell.
Panels (a) and (b) of Fig.\,\ref{bare MCD dec} show the left and right propagation within each sublattice as a function of time and the difference between the rightward propagation $\Delta x_R(t)= \langle x_R,A\rangle-\langle x_R,B\rangle$ on sublattices \(A\) and \(B\) for \(v/w = 1.01\), respectively. In panels (c) and (d) are reported the corresponding quantities for \(v/w = 0.99\). 
In Figs. \ref{omc8 dec}(a) (\ref{omc8 dec}(b)) and \ref{omc8 dec}(c) (\ref{omc8 dec}(d)) we plot the decomposed components of the MCD and the corresponding differences in the rightward propagation for $\alpha=0.1$ ($\alpha=0.3$). In this case the symmetry of the sublattice is broken by the interaction with the cavity modes.
As stressed previously, when the cavity frequency \(\omega_0\) is larger than the electronic energy scales \(v\) and \(w\), the effect of the interaction is essentially a renormalization of the intra-cell hopping.
For frequencies comparable to the electronic hopping amplitudes, the dynamics are inevitably slowed down due to dissipative effects, whose impacts becomes more pronounced as the interaction increases. Nevertheless, the coupling is still capable of inducing opposite polarizations on the left and right parts of the chain, while the overall polarization remains zero. 
Panels (a) and (b) of Fig.\,\ref{omc1 dec} show again, for a cavity frequency \(\omega_0 = 1.0w\) and \(v/w = 1.05\) in the weak-coupling regime, the decomposed components of the MCD and the sublattice difference in the rightward propagation, above a critical value of the coupling. By contrast, panels (c) and (d) refer to the strong coupling regime. One can notice how the propagation shown in Fig.\,\ref{omc1 dec}(c) is slower than in Fig.\,\ref{omc1 dec}(a). It takes a longer time scale to reach the asymptotic value. This behaviour is associated with the requirement of many bosonic oscillations in order to clearly reveal the effects of the electron–phonon interaction. Nevertheless, the interaction is able again to break the sublattice symmetry, signaling the induced topological properties.

\section{Conclusion}
In this work, we have investigated whether interaction-induced topological phenomena, previously identified at half filling in static many-body observables, leave detectable signatures also by exploting single-particle dynamics in an open SSH chain. Using tensor-network simulations, we have shown that coupling the SSH model to local cavity modes can produce clear dynamical markers of topology out of equilibrium, even when the system is initialized with a single particle located in the bulk or at the boundary.

In the antiadiabatic regime in which the cavity frequency is much larger than the electronic hopping amplitudes, the boson–fermion coupling effectively renormalizes the model parameters. This mechanism produces a discontinuous jump in the Mean Chiral Displacement and a finite survival edge probability. Both effects reveal that the cavity-mediated interaction acts as a genuine driver of topological transitions, consistent with previously reported many-body results.

When the cavity frequency becomes comparable to the electronic hopping amplitudes, the dynamics enter a competing regime with strong dissipative effects and slows down propagation while still allowing topological signatures to emerge. In this crossover regime, the MCD displays a smooth behavior but retains clear information on the induced topology.
This provides experimentally accessible signatures of cavity-induced topological phase transitions. 

Although numerical complexity remains a major limitation, several extensions of this study are possible. A natural generalization involves increasing the number of modes in the engineered baths. We expect that, for modes with frequencies beyond the bandwidth, the overall behavior of the MCD will remain essentially unchanged. However, the dependence of the survival edge probability on the characteristic frequency, which tends to increase at higher frequencies, may be strongly influenced by the detailed structure of the baths. Future work will therefore include an in-depth analysis of modes with frequencies comparable to the bandwidth, which are particularly challenging to handle because they can induce intraband processes and generate significant entanglement, making numerical simulations more demanding. Indeed, non-Hermitian effects in the electronic subsector induced by such coupling can, in principle, be exploited. Finally, extension of the model can be made to caso of particles with spin degrees of freedom \cite{perroni4} and to more realistic topological nanowires \cite{perroni5}.

\color{black}

\section*{Acknowledgements}
G.D.F. acknowledges financial support from PNRR MUR Project No. PE0000023-NQSTI. G.D.B. and C.A.P. acknowledge funding from IQARO (Spin-orbitronic Quantum Bits in Reconfigurable 2DOxides) project of the European Union’s Horizon Europe research and innovation programme under grant agreement n. 101115190. G.D.B., G.D.F. and C.A.P. acknowledge funding from the PRIN 2022 project 2022FLSPAJ ``Taming Noisy Quantum Dynamics'' (TANQU). C.A.P. acknowledges funding from the PRIN 2022 PNRR project P2022SB73K ``Superconductivity in KTaO3 Oxide-2DEG NAnodevices for Topological quantum Applications'' (SONATA) financed by the European Union - Next Generation EU. 

\bibliographystyle{apsrev4-2}
\bibliography{biblio}

\appendix
\section{Survival edge probability for the bare SSH chain}\label{appendix probability}
Consider the semi-infinite SSH Hamiltonian with the left boundary at site $A_1$:
\begin{equation}
H=\sum_{n\ge 1}\big(v\ket{A_n}\bra{B_n}+w\ket{A_{n+1}}\bra{B_n}+\text{h.c.}\big),
\end{equation}
with intra-cell hopping $v$ and inter-cell hopping $w$. 
We look for the eigenstates $\ket{k}$ in the continuum such that $H\ket{k}=E_k\ket{k}$. The eigenvectors satisfy:

\begin{subequations}\label{mode eqs}
\begin{align}
E_k \braket{A_n|k} &= v\braket{B_n|k} + w\braket{B_{n-1}|k}, \label{ev 1}\\
E_k \braket{B_n|k} &= v\braket{A_n|k} + w\braket{A_{n+1}|k}. \label{ev 2}
\end{align}
\end{subequations}

Since we are interested only in the propagation from the left edge, we have to impose a node to the left of the chain. In order to do this, we consider the following ansatz:
\begin{equation}\label{ansatz}
\braket{B_n|k}=\mathcal{N} \sin({kn}) ,
\end{equation}
with $k \in (0,\pi)$ and $\mathcal{N}$ a normalization factor. In this way we have that the amplitude at $n=0$ is vanishing for every $k$.
Inserting Eq.\,\eqref{ansatz} into Eq.\,\eqref{ev 1} we obtain:
\begin{equation}\label{A_n(k)}
    \braket{A_n|k}=\frac{\mathcal{N}}{E_k}\{ v \sin{(kn)} +w\sin[k(n-1)] \},
\end{equation}
and again inserting Eq.\,\eqref{A_n(k)} into Eq.\,\eqref{ev 2} yields 
\begin{multline}\label{Bnk expand}
    E_k \braket{B_n|k}= \frac{\mathcal{N}}{E_k}v \biggl\{ v \sin{(kn)}+w \sin[k(n-1)] \biggl\}+ \\ \frac{\mathcal{N}}{E_k}w \biggl\{ v\sin{[k(n+1)]}+w \sin(kn) \biggl\}= \\
    \frac{\mathcal{N}}{E_k}\biggl[(v^2+w^2)\sin(kn)+vw\bigg\{\sin[k(n-1)]+\sin[k(n+1)] \bigg\} \biggl]= \\
    \frac{\mathcal{N}}{E_k}\bigg[v^2+w^2+2vw\cos k \bigg]\sin(kn),
\end{multline}
where in the last equality we used the identity $\sin [k(n+1)]+\sin [k(n-1)]=2 \sin(kn)\cos k$.

By imposing consistency between Eq.\,\eqref{ansatz} and Eq.\,\eqref{Bnk expand} we find, as expected, $E_k^2=v^2+w^2+2vw\cos k$, which validates our choice of ansatz. The eigenstate can thus be written as

\begin{multline}\label{k ket}
    \ket{k}=\mathcal{N} \sum_{n \geq 1}\bigg\{ \frac{1}{E_k}[v\sin(kn)+w\sin(k(n-1))]\ket{A_n} \\ +\sin(kn)\ket{B_n} \bigg\}.
\end{multline}
Imposing the normalization condition $\braket{k|k'}=\delta(k-k')$ fixes the prefactor to $\mathcal{N}=\sqrt{2/\pi}$.

The open boundary condition ($B_0=0$) therefore produces continuum eigenstates that are standing waves
with a node at $n=0$. Their local weight on $A_1$ is
\begin{equation}\label{k coeff}
\big|\langle A_1|k\rangle\big|^2=\frac{2}{\pi}\,
\frac{v^2\sin^2 k}{E^2_k}=\frac{2}{\pi}\,
\frac{v^2\sin^2 k}{v^2+w^2+2vw\cos k}\, ,
\end{equation}
with $k\in(0,\pi)$. 

Now let us evaluate the zero-energy edge mode $\ket{\psi_L}$ which appears in the topological phase. 

To this end, we set $E_k=0$ in Eqs.\,\eqref{mode eqs}:
\begin{subequations}\label{zero mode equations}
\begin{align}
 v\braket{B_n|\psi_L} + w\braket{B_{n-1}|\psi_L}=0, \label{zero mode a} \\
 v\braket{A_n|\psi_L} + w\braket{A_{n+1}|\psi_L}=0. \label{zero mode b}
\end{align}
\end{subequations}
Using the node condition $\braket{B_0|\psi_L}=0$ in Eq.\,\eqref{zero mode a} we get $\braket{B_n|\psi_L}=0$ for every $n$.
On the contrary, from Eq.\,\eqref{zero mode b} we get $ \braket{A_{n+1}|\psi_L}=-(v/w)\braket{A_{n}|\psi_L}$, which implies:
\begin{equation}
    \braket{A_{n}|\psi_L}=\bigg(-\frac{v}{w}\bigg)^{n-1}\braket{A_{1}|\psi_L}.
\end{equation}
Expanding $\ket{\psi_L}$ in the basis of the sublattice A $\{\ket{A_n} \}$ and requiring the normalization condition for the edge state $\ket{\psi_L}$ we end up with the following condition:
\begin{equation}\begin{split}
    \sum_{n \geq 1} | \braket{A_n|\psi_L}|^2=| \braket{A_1|\psi_L}|^2 \sum_{n \geq 0} \bigg( \frac{v^2}{w^2}\bigg)^n= \\
    |\braket{A_1|\psi_L}|^2\frac{1}{1-v^2/w^2}=1,
\end{split}\end{equation}

which is finite if and only if $w>v$.
For $v<w$ there exists a normalizable zero mode on sublattice $A$:
\begin{equation}
|\psi_L\rangle=\sqrt{1-\Big(\frac{v}{w}\Big)^2}\;
\sum_{n=1}^{\infty}\!\left(-\frac{v}{w}\right)^{n-1}|A_n\rangle,
\end{equation}
whose overlap with $|A_1\rangle$ gives the edge weight:
\begin{equation}\label{edge weight}
   \big|\langle A_1|\psi_L\rangle\big|^2= 1-\bigg(\frac{v}{w}\bigg)^2.
\end{equation}

Now we exploit the chiral symmetry $\Gamma H \Gamma=-H$. Its repeated action implies $\Gamma H^m \Gamma=(-1)^m H^m$ for every integer $m$. As a consequence, the overlap evaluated on odd powers of the Hamiltonian has to vanish:
\begin{equation}
\begin{split}
\braket{A_1|H^{2m+1}|A_1}
= \braket{A_1|\Gamma H^{2m+1}\Gamma|A_1} \\
= -\braket{A_1|H^{2m+1}|A_1} = 0 .
\end{split}
\end{equation}

where we have used $\Gamma \ket{A_n}=\ket{A_n}$ and $\Gamma \ket{B_n}=-\ket{B_n}$. 

This allows us to retain only the even-power terms of the evolution operator:
\begin{equation}\label{evolution edge}
    \mathcal A(t)=\braket{A_1|e^{-iHt}|A_1}=\braket{A_1|\cos (Ht)|A_1}.
\end{equation}
 Finally, the completeness relation can be written as:
 \begin{equation}\label{completeness}
     \mathbb{1}=\Theta(w-v)\ket{\psi_L}\bra{\psi_L}+\int_{0}^{\pi} dk \ket{k}\bra{k},
 \end{equation}
where we have introduced the $\Theta(x)$ Heaviside function, that is $\Theta(x)=1$ if $x>0$ and vanishing otherwise.
Inserting Eq.\,\eqref{completeness} into Eq.\,\eqref{evolution edge}, we obtain $\mathcal{A}(t)$:

\begin{equation} \begin{split}
\mathcal A(t)=\langle A_1|e^{-iHt}|A_1\rangle
= \,&\Theta(w-v)\big|\langle A_1|\psi_L\rangle\big|^2
\\ &+\int_0^\pi\! \mathrm dk\;\big|\langle A_1|k\rangle\big|^2\cos\!\big(E_k\,t\big),
\end{split}\end{equation}
which, upon using the edge weight in Eq.\,\eqref{edge weight} and the coefficients in Eq.\,\eqref{k coeff}, takes the final form

\begin{equation}\begin{split}\label{final At}
\mathcal A(t)
=\Theta(w-v)&\Big(1-\frac{v^2}{w^2}\Big)
\\ &+\frac{2}{\pi}\int_0^\pi \mathrm dk\;
\frac{v^2\sin^2 k}{\,v^2+w^2+2vw\cos k\,}\;
\cos(E_k t),
\end{split}\end{equation}
where the first term is the discrete edge contribution present only for $v<w$.
The survival edge probability is then given by $p_1(t)=| \braket{A_1|e^{-iHt}|A_1}|^2$, and its long-time limit is:
\begin{equation}
\lim_{t\to\infty}p_{1}(t)=
\begin{cases}
\big(1-\frac{v^2}{w^2}\big)^2, & v<w,\\[2pt]
0, & v\ge w.
\end{cases} \:\: .
\end{equation}
\section{Numerical issues: why TDVP does not work?} \label{app TDVP}
The topology of the tensor network, i.e., the order in which the physical sites are arranged in the MPS chain, plays a crucial role in the computational efficiency.  
Two equivalent physical states may admit MPS representations with vastly different required bond dimensions.  
When two sites that are strongly entangled are placed far apart in the chain, their correlations must propagate through many intermediate bonds.  
Consequently, the local entanglement becomes distributed across several links, causing the bond dimension to grow in order to capture these long-range correlations.  
As a result, the bond dimension typically increases with the distance between highly correlated sites.  
In the following, we will see that the growth of the bond dimension during time evolution depends sensitively on the tensor ordering:  
in the most efficient configuration, it scales logarithmically with time, whereas in a poorly chosen ordering it grows linearly.  
This difference can lead to enormous variations in computational cost, even for identical physical parameters.

The Time-Dependent Variational Principle (TDVP) \cite{TDVP} provides a powerful framework for simulating the real-time evolution of quantum many-body systems within the MPS formalism. Two main implementations are commonly used: the one-site TDVP updates a single tensor at a time while keeping the bond dimension fixed, ensuring stable and strictly variational dynamics but limiting the growth of entanglement; the two-site TDVP, on the other hand, evolves two tensors simultaneously, allowing the bond dimension to adapt dynamically and thus capture the increase of entanglement during time evolution. However, this flexibility comes at the cost of a loss of strict unitarity.
In our case, the two-site TDVP algorithm fails to produce a stable time evolution due to the presence of three-body interaction terms in the Hamiltonian:  
\begin{equation} 
    H_{SSH-B}=g\sum_n (b^\dagger_n+b_n) H_{hop,n}.
\end{equation}
In Eq.\,\eqref{interaction hamiltonian}, $H_{hop_n}$ describes either the intra- ($c^\dagger_{n,A} c_{n,B}+h.c.$) or the inter-cell ($c^\dagger_{n+1,A} c_{n,B}+h.c.$) fermionic hopping.

The two-site TDVP constructs the tangent-space projector only for pairs of neighboring sites, and therefore cannot accurately capture operators acting on three sites. As a consequence, the projected time evolution becomes inconsistent with the exact Schrödinger dynamics, leading to numerical instabilities and unphysical behavior. This issue reflects a fundamental inadequacy of the local structure of the TDVP projection to capture the effects of extended range of the Hamiltonian interactions.  
For these reasons, when we apply TDVP (either one-site or two-site) to our specific model, the algorithm reproduces only the non-interacting case, regardless of the interaction strength, and fails to capture the boson–fermion coupling. Further attempts have been performed by following the GSE strategy proposed in Ref.\,\cite{gse}. Along this route the simulations became interaction dependent, but the large number of parameters made it very challenging to achieve clean convergence. 
To overcome these limitations, we instead perform simulations using tensor-network methods \cite{Scholl-Tensor-RevModPhys.77.259,TensorNetwork_Bridgeman_2017,RevModPhys.93.045003_Tensor,ORUS-tensornet}, implemented with the ITensor library in \textit{Julia} \cite{fishman2022itensor} and a package that provides the $W^I$ time-evolution algorithm, which is suitable for products of three operators \cite{ITensorMPS_issue101_2024} (unlike the previous C++ implementation).
\subsection{The optimal choice for the tensor chain}\label{convergence section}
  
\begin{figure}[htbp!]
    \includegraphics[width=0.9\linewidth]{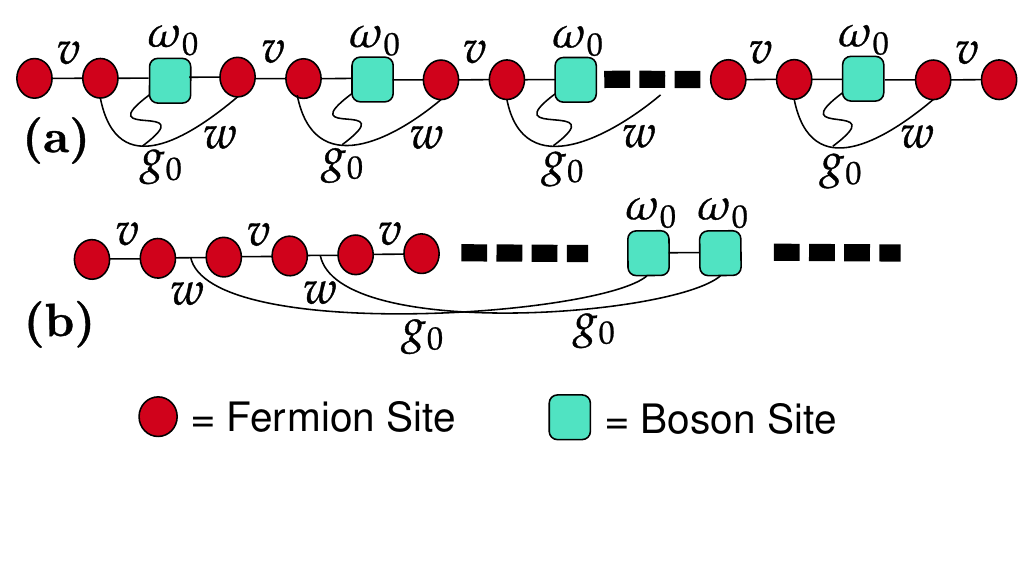}
    \caption{Two possible tensor orderings for our model, in the case of boson-fermion coupling through inter-cell hopping.}
    \label{tensor ordering}
\end{figure}

A given quantum state can admit different, physically equivalent MPS representations, some of which are far more numerically demanding than others. For clarity and concreteness, in our model two possible tensor orderings are shown in Fig.\,\ref{tensor ordering}.  

\begin{figure}[htbp!]
    \centering
\includegraphics[width=1.05\linewidth]{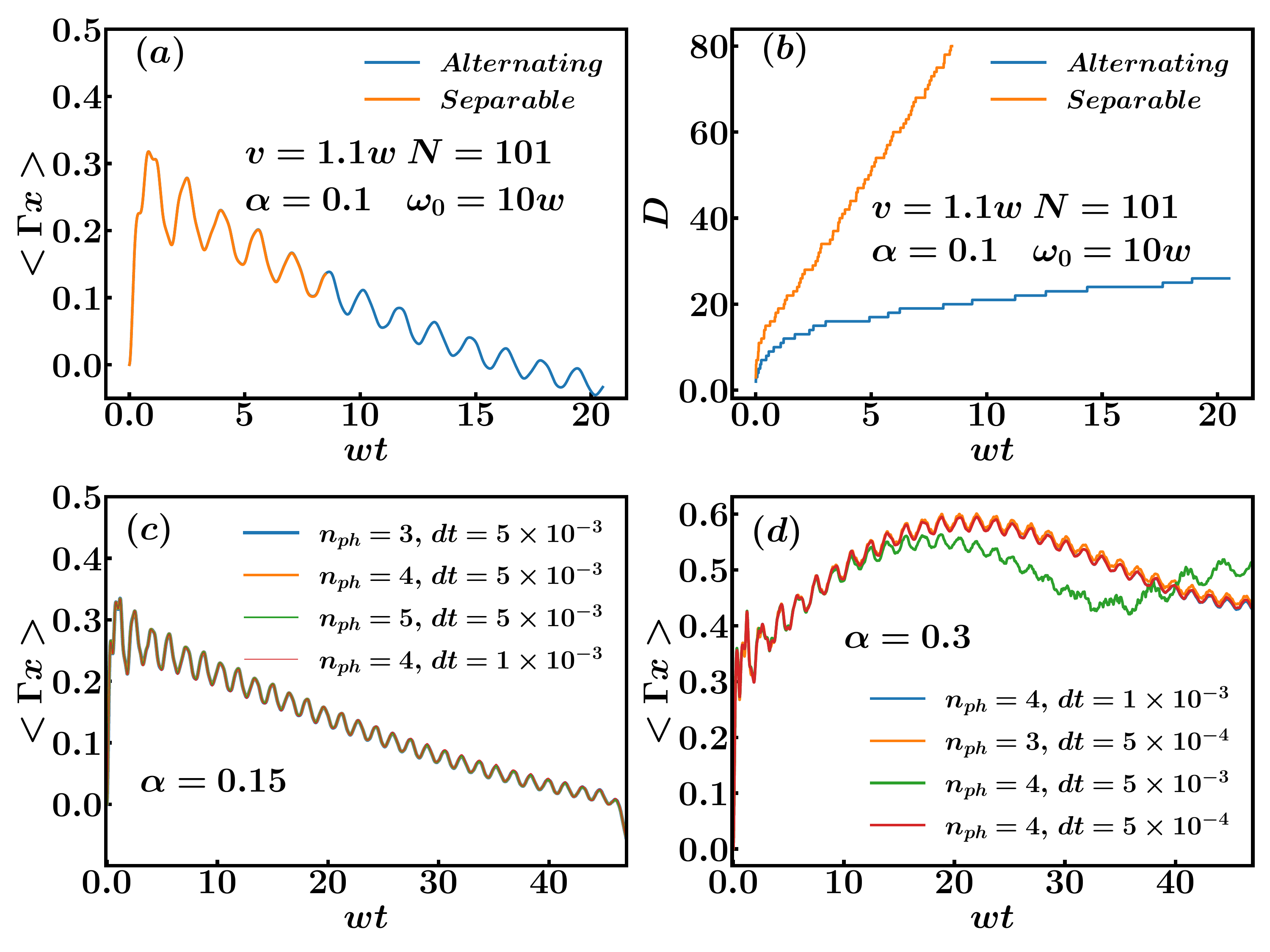}
    \caption{Panels (a) MCD and (b) growth of the bond dimension vs. time for the two different tensor configurations of Fig.\,\ref{tensor ordering}. Panels (c) and (d) show examples of convergence in the weak- and strong- coupling regimes, respectively.}
    \label{convergence MCD}
\end{figure}

In panel (a), our model alternates two fermionic sites (i.e., a unit cell) and one bosonic site, while panel (b) shows a topology with separated fermionic and bosonic parts.  
Version (a) is currently implemented in the numerical code and is the configuration that reproduces the results discussed in the main text. The reason is shown in Fig.\,\ref{convergence MCD}. 
For fixed physical parameters, from panel (a) we see that the two configurations reproduce the same MCD during the early-time dynamics, i.e. for $wt<10$, but already in this regime the difference in bond dimension $D$ is significant, as illustrated by \ref{convergence MCD}. Using the tensor ordering of panel (b) in Fig.\,\ref{tensor ordering} makes it impractical to perform simulations that reach the long-time limit. Naturally, in both configurations we must also fix the maximum number of bosonic excitations $n_{ph}$, which represents a further convergence parameter. Panels (c) and (d) show examples of convergence when varying the time step $dt$ and the cutoff $n_{ph}$ in the weak- and strong-coupling regimes, respectively. In particular, one can see that the convergence with respect to the time step is tricky, especially in the strong-coupling regime (see the green curve in panel (b) of Fig.\,\ref{convergence MCD}).

\section{Cavity modes coupled to the intra-cell hopping}
\begin{figure}[htbp!]
    \centering
    \includegraphics[width=0.9\linewidth]{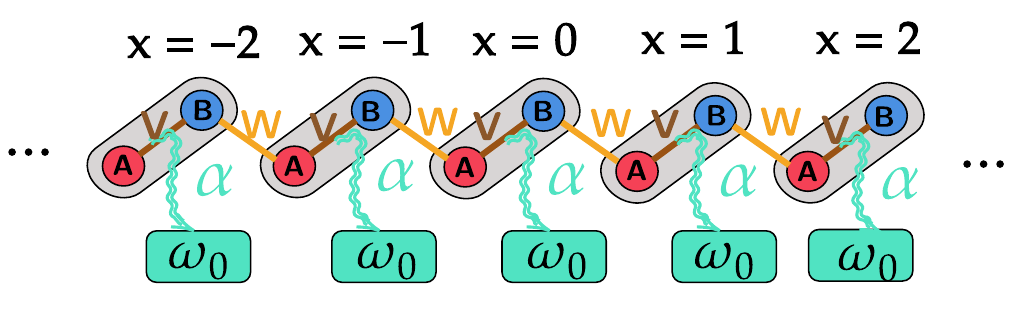}
    \caption{Open SSH chain with cavity modes coupled to the intra-cell hopping.}
    \label{fig13}
\end{figure}
\begin{figure}[htbp!] 
    \begin{center}   
        \includegraphics[scale=0.20]{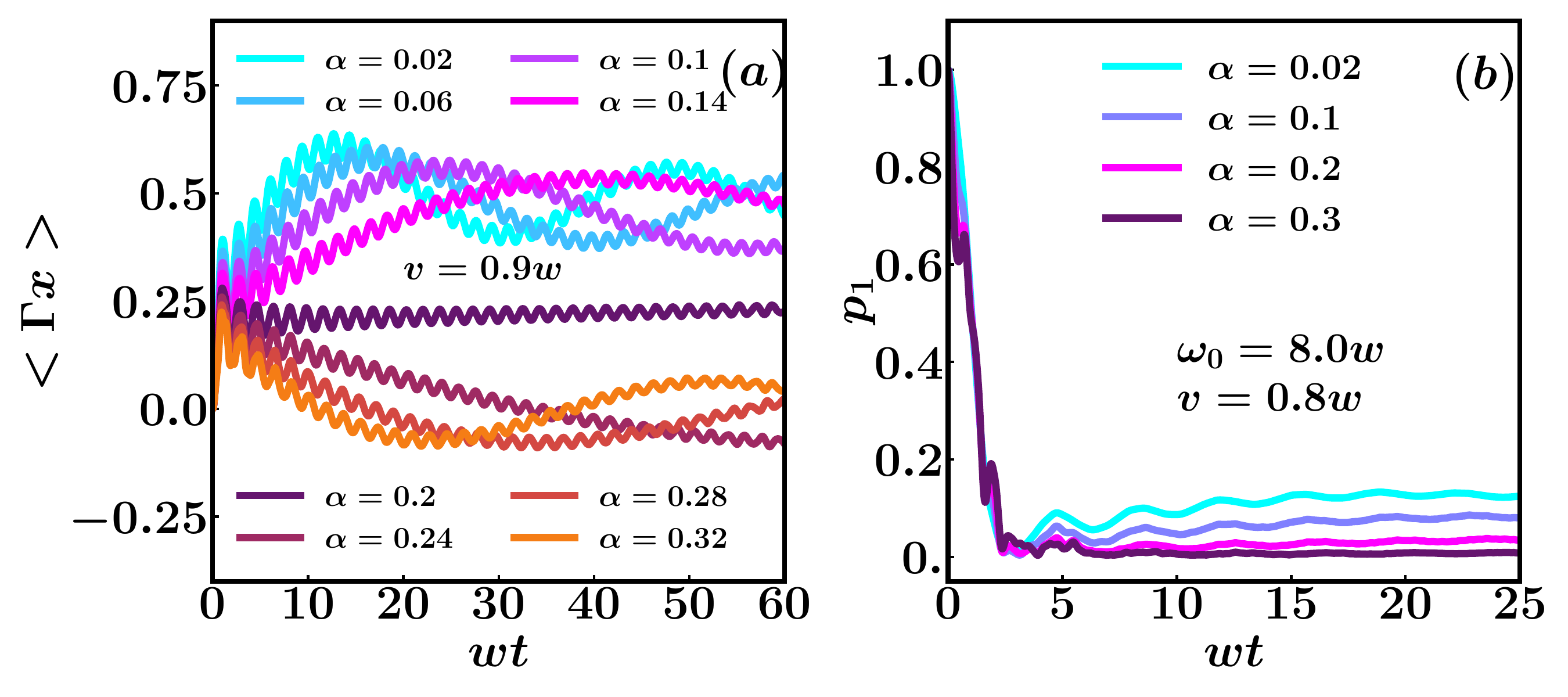} \caption{\label{fig14} Mean Chiral Displacement (panel (a)) and survival edge probability (panel (b)) as functions of dimensionless time, for different values of the coupling $\alpha$ and for $N=201$ unit cells.}
    \end{center} 
\end{figure}

For the sake of completeness, considering the cavity modes coupled to the intra-cell hopping $v$ (see Fig.\,\ref{fig13}), the opposite behavior can be observed. Starting from the condition $v<w$, the interaction drives the MCD to oscillate between a non trivial and a trivial topological invariant. This effect is further highlighted by the edge dynamics, which shows that the interaction can suppress such a survival probability. The results are reported in Fig.\,\ref{fig14}.

\newpage

\end{document}